\documentclass[journal]{IEEEtran}
\usepackage[noadjust]{cite}

\usepackage{graphicx}
\usepackage{tabularx,booktabs}
\usepackage{dingbat}
\usepackage{diagbox}
\newcolumntype{C}{>{\centering\arraybackslash}X}
\usepackage{hhline} 
\usepackage{cite}
\usepackage{amssymb}
\usepackage{graphicx}
\usepackage{multirow}
\usepackage[table, dvipsnames]{xcolor}
\usepackage{color}
\usepackage{pifont}
\usepackage{booktabs}

\usepackage[normalem]{ulem}

\newcommand*\rot{\rotatebox{90}}

\definecolor{fancyBlue}{RGB}{119,158,203}

\usepackage{scalerel}
\usepackage{tikz}
\usetikzlibrary{svg.path}

\definecolor{orcidlogocol}{HTML}{A6CE39}
\tikzset{
  orcidlogo/.pic={
    \fill[orcidlogocol] svg{M256,128c0,70.7-57.3,128-128,128C57.3,256,0,198.7,0,128C0,57.3,57.3,0,128,0C198.7,0,256,57.3,256,128z};
    \fill[white] svg{M86.3,186.2H70.9V79.1h15.4v48.4V186.2z}
                 svg{M108.9,79.1h41.6c39.6,0,57,28.3,57,53.6c0,27.5-21.5,53.6-56.8,53.6h-41.8V79.1z M124.3,172.4h24.5c34.9,0,42.9-26.5,42.9-39.7c0-21.5-13.7-39.7-43.7-39.7h-23.7V172.4z}
                 svg{M88.7,56.8c0,5.5-4.5,10.1-10.1,10.1c-5.6,0-10.1-4.6-10.1-10.1c0-5.6,4.5-10.1,10.1-10.1C84.2,46.7,88.7,51.3,88.7,56.8z};
  }
}

\newcommand\orcidicon[1]{\href{https://orcid.org/#1}{\mbox{\scalerel*{
\begin{tikzpicture}[yscale=-1,transform shape]
\pic{orcidlogo};
\end{tikzpicture}
}{|}}}}

\usepackage{hyperref}

\usepackage[colorinlistoftodos]{todonotes}   

\usepackage{amsmath}
\usepackage{algorithmic}

\usepackage{array}
\usepackage{url}

\usepackage{xcolor}

\begin{document}
%
\title{Applications of Mechanism Design in Market-Based Demand-Side Management}
%
%
%

\author{Khaled~Abedrabboh\textsuperscript{\orcidicon{0000-0001-7572-0158}}
        ~and~Luluwah~Al-Fagih\textsuperscript{\orcidicon{0000-0002-0449-1324}}
\thanks{The authors are with the College of Science and Engineering, Hamad Bin Khalifa University, Doha 34110, Qatar e-mail: lalfagih@hbku.edu.qa }
\thanks{L. Al-Fagih is with Faculty of Science, Engineering and Computing, Kingston University London, Kingston upon Thames KT12EE, U.K.}
}

\maketitle

\begin{abstract}
The intermittent nature of renewable energy resources creates extra challenges in the operation and control of the electricity grid. Demand flexibility markets can help in dealing with these challenges by introducing incentives for customers to modify their demand. Market-based demand-side management (DSM) have garnered serious attention lately due to its promising capability of maintaining the balance between supply and demand, while also keeping customer satisfaction at its highest levels. Many researchers have proposed using concepts from mechanism design theory in their approaches to market-based DSM. In this work, we provide a review of the advances in market-based DSM using mechanism design. We provide a categorisation of the reviewed literature and evaluate the strengths and weaknesses of each design criteria. We also study the utility function formulations used in the reviewed literature and provide a critique of the proposed indirect mechanisms. We show that despite the extensiveness of the literature on this subject, there remains concerns and challenges that should be addressed for the realistic implementation of such DSM approaches. We draw conclusions from our review and discuss possible future research directions. 
\end{abstract}


%
\IEEEpeerreviewmaketitle


\section{Introduction}
\label{intro}
%
%
%
%


\IEEEPARstart{T}{he} global energy demand is expected to grow significantly in the near future \cite{IEA2020}. A move towards a larger electricity share of that demand is envisaged in order to limit climate change and achieve sustainable development. A large portion of this growth in demand is expected to be met by renewable energy resources, from both large scale generation and small scale distributed generation. However, due to the uncertainty and intermittency of renewable energy resources, this will prompt additional challenges in maintaining the balance between supply and demand \cite{IMPRAM2020100539}.  Also, to further ensure the reliability and security of supply, utilisation of the electricity grid infrastructure should be enhanced. This can also help in limiting the investments required to improve the grid infrastructure. These challenges have caused significant interest in promising demand participation techniques. 

There have been many technological developments that can allow bidirectional power and communication flow between grid entities. Combined, these technologies can empower the transition towards a smart grid, permitting all grid portions to participate in the management of energy flow. 
The design of a comprehensive demand side management (DSM) scheme is an essential part of the smart grid. DSM refers to the methods used to adjust the load profile of an electricity grid in a way that profits both supply and demand \cite{SHARIFI2017review}. Its main benefits include maintaining the balance between supply and demand \cite{Balijepalli2011} and deferring some of the required investments in electricity infrastructure \cite{ALBADI2008}. However, for these schemes to successfully engage prosumers (\textit{pro}ducer and con\textit{sumer}) in abandoning some of their comfort and providing flexibility to the grid, the design of an attractive market structure that is beneficial to all stakeholders is paramount \cite{chrysikou2015review}.

\begin{figure}[h]
\centering
\includegraphics[width=0.45\textwidth]{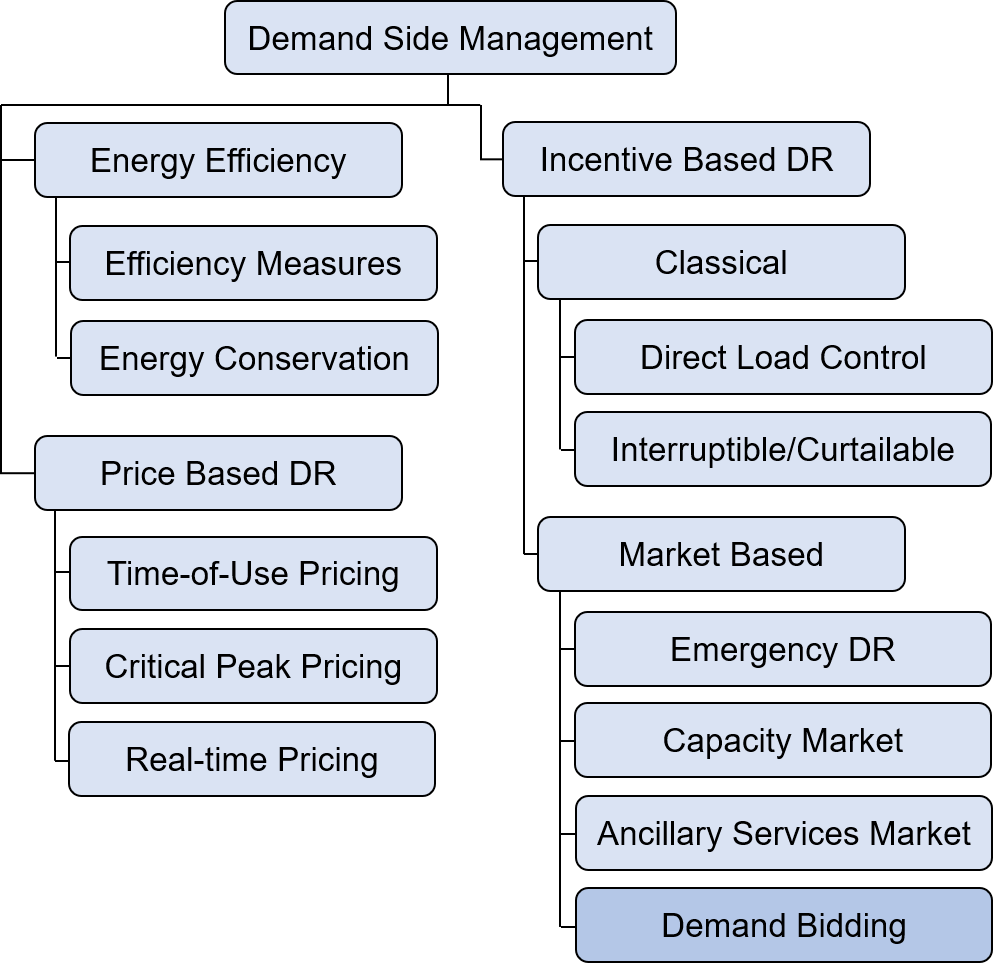}
\caption{Categorisation of DSM methods, highlighting the scope of this review.}
\label{fig: Fig 2}
\end{figure}

DSM schemes can be categorised into energy efficiency and demand response (DR), which can refer to either price-based demand response (PBDR) or incentive-based demand response (IBDR) (\cite{ALBADI2008, Palenski2011, YAN2018}). Fig.~\ref{fig: Fig 2}, shows the different methods used for DSM and their classification. Energy efficiency can either refer to improving the efficiency of demand, such as using insulation in buildings in order to reduce the cooling/heating demand, or it can refer to changing consumption behaviour to a more efficient one, such as refining the thermostat temperature to reduce power consumption. This concept is often referred to as energy conservation \cite{Boshell2008}.
PBDR techniques aim to flatten the demand profile by moving from static (demand-independent) electricity rates to more dynamic pricing methods. These methods include Time of Use (ToU) pricing, Critical Peak Pricing (CPP) and Real-Time Pricing (RTP). Even though these techniques have shown positive results in demand peak shifting (\cite{YAN2018, SRINIVASAN2017132}), they do not take the consumers' preferences into consideration, and as a result, customers' responsiveness to such dynamic pricing techniques can be limited (\cite{YAN2018, Reiss2005}). On the other hand, IBDR have provisions for customers' individual preferences \cite{Zhang2012review}. Its programmes provide rewards for customers as an encouragement to voluntarily participate in DR. IBDR methods can be further categorised into classical and market-based methods. In classical IBDR, customers are contracted to either allow control of their load to the system operator, referred to as direct load control (DLC), or curtail their demand at an announced DR event, referred to as interruptible/curtailable (I/C). \cite{Khajavi2011review}. Alternatively in market-based IBDR, customers are rewarded if they choose to participate in a DR event \cite{ALBADI2008}. Emergency DR, for instance, offers performance-based incentives to customers who voluntarily curtail their demand in contingencies \cite{Khajavi2011review}. In capacity market programmes, participants can choose to commit to undertake a specified amount of demand reduction when called upon \cite{ALBADI2008}. Ancillary services market programmes are often offered to large consumers or demand response aggregators (DRA) who can bid their load curtailment services in the spot market. Accepted bids are then put on standby and participants who are called upon need to provide their load curtailment services at a short notice \cite{Khajavi2011review}. Demand bidding is arguably the most promising method in DR as it allows customers to actively participate in the electricity market \cite{Zhang2012review}. Here, customers can individually or cooperatively bid for either their total demand schedule or their offered demand variation to the system operator \cite{PATERAKIS2017871}. Demand bidding can increase the price elasticity of demand and revolutionise the electricity market, especially when distributed energy resources such as distributed RE and electricity storage systems are used \cite{Zhang2012review}.

Many DSM models are based on game theoretic approaches (\cite{chrysikou2015review, ESTHER2016survey}). These build on the fact that consumers (and prosumers) act rationally and selfishly, i.e.~make logical decisions that are in their best self-interest. Indeed, game theory can help in analysing such behaviour in competitive public resource markets, such as transportation \cite{Amar2018GT}, healthcare \cite{Abedrabboh2021GT} and energy (\cite{pilz2017energy, pilz2019dynamic}). Mechanism design (MD) is the normative part of game theory, where the rules (mechanism) of the market are not given, but designed to achieve a certain objective \cite{Fudenberg}. Indeed, researchers have drawn parallels between the setting of an MD problem and that of DSM. Its ability to achieve social optimal outcomes in auctions and public good markets has given rise to the application of MD in DSM schemes, especially demand bidding IBDR \cite{Li2020MDreview}. Although there are several recent reviews available in the literature on game theoretic DSM (\cite{Pilz2019review, SHARIFI2017review, Loni2017review}), none are dedicated to MD approaches, despite the extensive research that has been done in this field.

In this paper, we review MD models that are proposed for demand bidding IBDR. We present different criteria for the classification of such models and discuss the advantages/disadvantages of each class. We also review the formulations and assumptions pertaining to the utility function used in the proposed mechanisms. Furthermore, we discuss the potential of indirect mechanisms in DSM schemes and provide a critique of the ones proposed in the literature. Based on this review, we draw conclusions and discuss potential future research directions. Our contributions are threefold:
\subsubsection{classification of MD applications in DSM}
We classify the available literature on DSM schemes that adopt MD in their methods based on four criteria, \textit{revelation, allocation, sequence} and \textit{scalability}. We discuss the importance of the used criteria and analyse the properties and strengths/weaknesses of each design category. The classification of MD-DSM schemes can provide insights into the areas that are not thoroughly investigated yet and that would require further research.

\subsubsection{survey of utility functions}
We provide an in-depth examination of the utility function types used in the existing literature that is within the scope of this review. To the best of our knowledge, this is the first time a detailed survey of the utility function for energy is provided. Three types of utility functions were found in the reviewed papers. The limitations of these types are discussed in detail.

\subsubsection{review of indirect mechanisms}
A promising subfield of mechanism design applications in DSM is the employment of \textit{indirect} mechanisms. 
A comprehensive review of DSM models that adopt such mechanisms is provided, where we discuss the concerns that are associated with such mechanisms.

To be able to understand the concepts and definitions used throughout this review, the basic principles of MD are presented in Section~\ref{sec:MD}. In Section~\ref{sec:class}, we classify the MD mechanisms proposed for DSM and provide a critique of each category. The types of utility functions used in these mechanisms are then investigated in Section~\ref{sec:utility}. Section~\ref{sec:indirect} provides a thorough review of the indirect mechanisms that are proposed for DSM. Conclusions and future research directions are discussed in Section~\ref{sec:con}.


\section{Background on Mechanism Design}
\label{sec:MD}

Mechanism design is the normative part of game theory. It is relevant (applicable) when a \textit{principal} is assigned with setting the rules of the market so that a desired economic objective can be achieved \cite{Fudenberg}. One of the common applications of MD is the pricing and provision of public goods. Given the opportunity, consumers acting in their own self-interest would exploit the public resource. This is known as the \textit{tragedy of the commons} \cite{Hardin}. To avoid this social dilemma, the mechanism designer (i.e.~principal) wishes to choose a socially preferred outcome (allocations of the public good) that guarantees the best social economic welfare. This can be a difficult task considering that the mechanism designer does not know the preferences of the market participants and how much they value this public good. Another example of this is an auction, where the principal (auctioneer) does not know the participants' private valuation of the auctioned items. Nonetheless, the auctioneer's objective is to allocate the auctioned goods to the bidders who value them the most. Due to the preferences being privately known to the market participants and unknown to the mechanism designer, MD has a similar setting to that of Bayesian games, in which players have incomplete information about other players. In order to understand the tools that are available to the mechanism designer, we first need to analyse the setting of a Bayesian game \cite{jackson2014mechanism}, which has:

	\begin{enumerate}
	\item A set of \textit{agents} (i.e.~players) $\mathcal{N} = \{1,\dots,n\}$, where $n$ is the total number of market participants. All agents are assumed to be selfish and rational, i.e.~utility maximising, where utility is the benefit they get from consuming a product or hiring a service.
	\item A set of possible joint \textit{type} vectors, $\Theta = \Theta_1 \times \dots \times \Theta_n$. An agent's type describes their private valuation of a good or service, however it can be extended to include any information that is private and that can indicate their preferences.
	\item A set of \textit{outcomes}, $\Phi$. An outcome $\phi$ can include a set of allocations $\mathcal{Q}$ and a set of monetary transfers $\mathcal{T}$. It can represent the decisions made by the mechanism designer centrally or it can be the result of the actions made by the agents in a distributed manner.
	\item A set of utility functions $u=\{u_1, \dots, u_n\}$, where $u_i:\Phi \times \Theta \mapsto \mathrm{I\!R}$. A utility function is a mathematical representation of an agent's preferences and is a function of consumption quantity and their type. Therefore, $u_i (q_i , \theta_i)$ is the utility agent $i$ of type $\theta_i$ gets from consuming allocation $q_i$ as part of an outcome $\phi$. Utility also depends on the payment $t_i$ an agent makes for their consumption and their overall utility becomes $u_i (q_i , \theta_i) - t_i$.
	\end{enumerate}%

The mechanism designer's objective is to implement a social choice function which maps the preferences of all agents within the society to a certain outcome. However, given that these preferences are private information, the mechanism designer needs to elicit this information by defining the rules of the mechanism $\mathcal{M} = \left \langle \mathcal{B}, \pi, \mu \right \rangle$, which has: (1) a message (bid) space $\mathcal{B}$, representing the set of possible bids, (2) an allocation rule $\pi$ that maps the agents' preferences (bids) to an outcome $\pi : \mathcal{B} \mapsto \Phi$, and (3) a payment rule $\mu$ that dictates, based on the agents' bids, the amount of monetary transfer each agent has to make $\mu : \mathcal{B} \rightarrow \mathrm{I\!R}^N$ \cite{KRISHNA2010MD}. 

In order to understand the properties of different mechanisms, the following definitions are presented and briefly discussed. These definitions are reproduced from (\cite{nisan2007, KRISHNA2010MD, jackson2014mechanism, Narahari2014}).

\textbf{Definition 1: }\textit{Economic Efficiency}.
A mechanism is called \textit{efficient} if its allocation rule maximises the social welfare of its agents. Social welfare is the aggregate utilities of participating agents $\sum_i^N u_i$, or:

    \begin{equation}
		    q^\ast(\theta) = argmax\sum_{i\in\mathcal{N}}u_i\left ( q_i, \theta_i \right )
		    \label{eqn:SWO}
	\end{equation}


\textbf{Definition 2: }\textit{Direct Revelation}.
\label{sec:revelation}
The product allocations that guarantee maximum Social welfare can be determined by solving the optimisation problem in Eq.~\ref{eqn:SWO}. However, the mechanism designer does not have access to the private preferences that are essential to solving this problem.
In a direct revelation mechanism, the principal simply asks the agents to report their private information as part of their bid space. This simplifies the mechanism design problem, however, given the selfish nature of individual agents, it raises another concern that agents might be tempted to report their private types untruthfully. As false reporting would compromise the efficiency of the mechanism, it is imperative that agents are incentivised to report their types truthfully. This is known as the \textit{incentive compatibility} constraint (see definition below). In contrast, indirect mechanisms are privacy-preserving mechanisms that do not require agents to fully report their private information. This can be done by collecting incremental information from the agents in a multiround (i.e.~iterative) mechanism.

\textbf{Definition 3: }\textit{Incentive Compatibility}.
This constraint is essential to ensure \textit{truthfulness} in direct mechanisms and guarantee economic efficiency. To ensure that agents, acting in their own self-interest, report their private types in a truthful manner, the mechanism designer must assert that agents do not gain from false bidding. The incentive compatibility constraint is formulated as:
	\begin{equation}
		u_i (q_i, \theta_i) \geq u_i (q_i, \theta) ~~~\forall i \in \mathcal{N} 
		\label{eqn:IC}
	\end{equation}
This means that the utility $u_i (q_i, \theta_i)$ agent $i$ gets from reporting their true type $\theta_i$ is never less than what they get from untruthful bidding $u_i (q_i, \theta)$.

\textbf{Definition 4: }\textit{Individual Rationality}.
Individual rationality (IR), or \textit{participation} constraint is a desired property of mechanisms in which participation is optional. It ensures that agents receive at least as much utility as they would get by not taking part. Assuming that an agent gets a payoff of zero by non-participation, IR constraint can be formulated as \cite{Narahari2014}:
	\begin{equation}
		u_i (q_i, \theta_i) \geq 0 ~~~\forall i \in \mathcal{N} 
		\label{eqn:IR}
	\end{equation}

\textbf{Definition 5: }\textit{Budget Balance}.
To prevent monetary losses from being incurred by the mechanism, the coordinating agent (i.e.~mechanism designer) should select a mechanism that guarantees budget balance, or $\sum_i^N t_i(\theta) \geq 0$. This means that the mechanism does not require external subsidies as it either generates money or preserve its balance. 

\section{classification}
\label{sec:class}


Mechanism design can help in selecting social outcomes that avoid inefficiencies in the allocation of a good or service. For this reason, many scientists have adopted MD techniques in their approaches to DSM. To the best of our knowledge, \cite{Samadi2011} is the first work that proposes an MD approach to DSM. Since then, the research on DSM using MD has been extensive. In order to have an overview of the different mechanisms proposed for DSM, we investigate four criteria, based on which we categorise all the proposed models. These four classes are discussed below. Table~\ref{tab:class} lists the proposed mechanisms in their respected categories.

\begin{table} 
	\centering
    \caption{Classification of reviewed mechanisms based on criteria described in section~\ref{sec:class}.}
    \label{tab:class}
    \begin{tabular}{|c|c|cc|cc|cc|cc|}
        
        \hline
        &&\multicolumn{2}{c}{revelation} \vline & \multicolumn{2}{c}{allocation} \vline & \multicolumn{2}{c}{sequence} \vline & \multicolumn{2}{c}{scalability} \vline
        \\
        \cline{3-10}
        
        \rot{\# papers}&ref&\rot{direct} & \rot{indirect} & \rot{central} & \rot{distributed } & \rot{iterative} & \rot{one-shot} & \rot{scalable} & \rot{unscalable }\\ \hline
	\rowcolor{fancyBlue!25}
	27
	&\cite{Samadi2011, Li2011, Cao2012, Samadi2012, Chen2014, Ma2014, Salfati2014, Mhanna2014towards, Xu2015challenges, Li2015, Hayakawa2015, Li2016marketpart1, Li2016marketpart2, bistarelli2016mechanism, Giuliodori2016AMD, Sinha2017, Tavafoghi2017, ma2017generalizing, Li2017reliability, tanaka2018, Li2018astrategy, 8Zhong2018auction, Chen2019operating, Borges2019, Mediwaththe2019, DAKHIL2019, Li2020towards_differential, Tsaousoglou2020mechanism}   & \checkmark      &          & \checkmark       &             &           & \checkmark        &          & \checkmark \rule{0pt}{2.6ex} \\
	21
	&\cite{Kota2012, Strohle2014, Mhanna2014, chau2014truthful, Xiang2015, Zhou2015demand, Zhou2015anonline, Chau2016truthful, AKASIADIS2017cooperative, Akasiadis2017mechanism, Yuan2017amechanism, meir2017contract, Zhong2017efficient, methenitis2019forecast, Khorasany2019, Afzaal2020agent, AlAshery2020, Roveto2020cooptimization, Sypatayev2020, wei2020mechanism, Zhong2020multiresource} & \checkmark      &          & \checkmark       &             &           & \checkmark        & \checkmark        &\rule{0pt}{2.6ex}\\
	\rowcolor{fancyBlue!25}
	3
	&\cite{Kang2017, Zhao2018towards, Zhong2018ADMM}  & \checkmark      &          & \checkmark       &             & \checkmark         &          & \checkmark        &\rule{0pt}{2.6ex}\\
	3
	&\cite{Okajima2014, xu2015incentive, MURAO2018real}  & \checkmark      &          &         & \checkmark           &           & \checkmark        &          & \checkmark\rule{0pt}{2.6ex}\\
	\rowcolor{fancyBlue!25}
	13
	&\cite{CHAKRABORTY2014, Jain2014, Chakraborty2015, Mhanna2016afaithful, Chapman2016algorithmic, STROHLE2016local, Ehsanfar2018, VUELVAS2018limiting, Vuelvas2018acontract, Aizenberg2019, Muthirayan2019, Jain2020amultiarmed, Muthirayan2020Mechanism} & \checkmark      &          &         & \checkmark           &           & \checkmark        & \checkmark        &\rule{0pt}{2.6ex}\\
	3
	&\cite{Faqiry2016doublesided, Mhanna2018afaithful, Zhou2018contract}  & \checkmark      &          &         & \checkmark           & \checkmark         &          & \checkmark        &\rule{0pt}{2.6ex}\\
	\rowcolor{fancyBlue!25}
	1
	&\cite{Majumder2015}  &        & \checkmark        & \checkmark       &             & \checkmark         &          &          & \checkmark\rule{0pt}{2.6ex}\\
    3
    &\cite{Faqiry2017abudget, Faqiry2019double, Faqiry2019distributed}  &        & \checkmark        & \checkmark       &             & \checkmark         &          & \checkmark        &\rule{0pt}{2.6ex}\\
	\rowcolor{fancyBlue!25}
	13
	&\cite{LiChen2011, Tushar2015, Namerikawa2015, Chapman2017aniterative, STERIOTIS2018anovel, Zhou2018designing, TSAOUSOGLOU2019Near, Tsaousoglou2019personalized, BARRETO2019, Hou2020reinforcement, Latifi2020arobust, Namerikawa2020distributed, Wang2020reward}  &        & \checkmark        &         & \checkmark           & \checkmark         &          & \checkmark        &\rule{0pt}{2.6ex}\\
	5
	&\cite{Bitar2013, Barreto2013, Barreto2015, Bitar2017, Tan2019posted}  &        & \checkmark        &         & \checkmark           &           & \checkmark        & \checkmark        &\rule{0pt}{2.6ex}\\
	\hline
    \end{tabular}
\end{table}

\subsubsection{Revelation}
One of the key factors in mechanism design is whether agents are required to report their private preferences directly and fully. These direct mechanisms are used to simplify the principal's task of implementing a social choice function. They can also guarantee the same level of efficiency (competitive ratio) as any other indirect mechanism. In fact, the \textit{revelation principle} (\cite{Gibbard1973, Myerson1979}) states that if a social choice function can be implemented by any mechanism, it can be implemented by a direct one without any loss of payoff. Nonetheless, direct mechanisms become less desirable in settings where agents prefer to keep their information private. This is a limitation when designing mechanisms for DSM as consumers may be reluctant to share their demand information and their value of that demand. Another limitation of direct mechanisms is that the communication overhead can become inefficient in markets that have a large number of agents, especially when multi-dimensional preferences are elicited. This concern can hinder the feasibility of direct mechanisms for DSM as most DSM applications need to engage a large number of consumers to ensure a certain level of demand flexibility. This communication overhead becomes aggravated as many DSM mechanisms ask agents to report their day-ahead consumption schedules and sometimes even ask for appliance-based demand information. Furthermore, computational complexity can be a concern in direct mechanisms because all the computation burden of solving the social welfare optimisation problem and determining the allocations/payments for each agent is incurred by the mechanism designer. This can limit the scalability of such mechanisms. 

Indirect mechanisms on the other hand are privacy preserving in the sense that agents are not required to share their information fully. Indeed, a social choice function can be implemented through an indirect mechanism by collecting sequential bids from agents without revealing their complete preferences. Although indirect mechanisms tend to be tractable as they can be implemented in a distributed fashion, they generally suffer from efficiency losses. A thorough review of the indirect mechanisms for DSM is provided in Section~\ref{sec:indirect}.

\subsubsection{Allocation}
Another key factor in the design of DSM mechanisms is the architecture used in their supply chain system. 
A central architecture refers to a supply chain system where a central entity (\textit{the principal}) makes allocation decisions based on its interactions with its agents. This can be inconvenient for consumers who are not equipped with distributed resources or storage systems, and although most of the mechanisms that adopt this architecture can optimise social welfare and yield efficient outcomes, adopting them in practice might result in significant discomfort for the consumers. Additionally, electricity consumers are usually unable to forecast their consumption accurately. Because of this, central allocation mechanisms are often coupled with reward/penalty incentives, which depend on how accurately consumers follow their assigned allocations. This uncertainty in demand does not only harm the efficiency of such mechanisms but also limit their acceptance among consumers. Alternatively, distributed mechanisms adopt an architecture where agents make their allocation decisions locally. In this architecture both the mechanism designer and the agents solve for the allocations and payments that optimise social welfare, thus sharing the computational burden of the mechanism. Although this architecture can ease the computation of DSM mechanisms and is more favorable to consumers than the central allocation architecture, it raises concerns about how faithfully agents choose and implement their allocation schedules.

\subsubsection{Sequence}
Some of the reviewed DSM mechanisms require agents to report their private preferences in one-shot. Others, in contrast use an iterative structure where agents update their bids at each step in response to a signal received from the mechanism designer. This is generally implemented to either ease the computational burden on the principal side (\cite{Faqiry2016doublesided, Mhanna2018afaithful, Zhou2018contract}) or to preserve the privacy of agents through indirect mechanisms (\cite{Tushar2015, Faqiry2019double, Faqiry2019distributed, Chapman2017aniterative, Namerikawa2020distributed}). Mechanisms that adopt the one-shot bidding process are more robust to communication delays. They can also be faster to implement. However, they are often less practical than iterative mechanisms as it is difficult for consumers to integrate their preferences into one bid, especially in a day-ahead setting as proposed in most DSM schemes. Nonetheless, although iterative mechanisms can overcome this limitation by asking agents to respond to their signals at each iteration, they usually have two main disadvantages; convergence in such schemes is generally slow, especially when a large number of participants need to transmit their bids at each iteration. This can be exacerbated when communication failures or delays are considered. The second limitation is that experienced agents that have an estimate of the number of iterations required for convergence can be untruthful in their bids at intermediary iterations if it generates higher utility gains. This can compromise the strategy proofness of such mechanisms.

\subsubsection{Scalability}
To maximise the benefits of DSM, broad participation of all types of demand is encouraged. Indeed, both large and small customers should engage in demand flexibility markets to ensure the success of such schemes \cite{PATERAKIS2017871}. Given that the number of electricity consumers is significantly large, scalability is one of the most important features of demand side participation mechanisms. Some works, (e.g.~\cite{Chapman2017aniterative, Mhanna2018afaithful, VUELVAS2018limiting, Faqiry2019distributed}), group a number of small residential customers under one agent, who is often referred to as a demand response aggregator (DRA). Thus proposing a two-level hierarchical structure for DSM where small customers and the DRA interact at the lower level, while DRAs offer their flexibility services to the grid at the higher level. Although this can significantly reduce the number of agents at the both grid and DRA levels, the number of agents at the lower level can still be quite large ($\sim1000$ \cite{Chapman2017aniterative}). Therefore, scalability is an essential requirement for the practical implementation of DSM mechanisms.

One of the most well known mechanisms that can ensure efficient allocation of goods/services is the \textit{Vickrey-Clarke-Groves} (VCG) mechanism. VCG is a direct and central mechanism, where agents report their private utilities to the principal who then determines their allocations by solving the social welfare optimisation problem. The payment rule in VCG is based on Clarke pivot rule, which requires that each agent pays their social cost of participation, which is the loss in welfare of all the other agents due to the participation of that agent. This payment rule ensures that the best strategy for agents who act selfishly is to report their private information truthfully. Many of the MD-DSM approaches are based on VCG (\cite{Samadi2011, Samadi2012, Xu2015challenges, Xiang2015, Zhou2015demand, bistarelli2016mechanism, Giuliodori2016AMD, meir2017contract, Zhong2017efficient, tanaka2018, Zhong2018ADMM, 8Zhong2018auction, Borges2019, methenitis2019forecast, Mediwaththe2019, AlAshery2020, Zhong2020multiresource, Okajima2014}). Although VCG can guarantee economic efficiency and incentive compatibility, it raises privacy and comfort concerns when applied to electricity market settings due to its direct revelation and central allocation structure. Additionally, it requires that agents' valuation functions (i.e.~utility functions) follow a strictly concave form. This does not only limit the representation of agents' preferences but also renders the social welfare optimisation problem intractable \cite{Li2011}. VCG is also vulnerable to collusion and shell bidding \cite{Yoram2010collusion}. Payments can also vary widely between agents and those with high valuations get higher allocations, which can be considered unfair, especially in settings where consumption is essential to preserve the quality of life such as electricity.

\section{the utility function}
\label{sec:utility}


Utility is a numerical measure of the satisfaction an individual gets from consuming a good or receiving a service. The utility function is a mathematical representation of this satisfaction as a function of the individual's preferences over a set of outcomes \cite{morgenstern1953theory}. In settings where transferable utility is possible through monetary payment, the utility function is said to be quasi-linear, i.e.~linear in payment, or reward. This linearity in monetary transfer is depicted in Eq.~\ref{eqn:Quasi},  where $U$ refers to the benefit a customer gets, which is represented by the difference between the customer's satisfaction and their payment. Alternatively, in the case that the agent is the seller or service provider, their utility can be formulated as the difference between the reward they get from providing that service and the associated cost of providing it. 

\begin{equation}
    \label{eqn:Quasi}
    U = u (q, \theta) - p 
\end{equation}

Representing the level of satisfaction of an agent as a function of consumption quantity and their preferences can be a difficult task. This is because preferences tend to be multi-dimensional. For instance, the utility of electricity consumption that is related to entertainment appliances (such as TV) is usually dependent on the time of consumption rather than quantity of consumption. Therefore, capturing these individual private preferences into one mathematical formulation that is suitable to all participating agents can be tricky. Nonetheless, the design of most mechanisms depend on making some underlying assumptions about the utility function. Additionally, to simplify the computation of their outcome and reduce their communication overhead, some direct mechanisms rely on a closed formulation of the utility function that can reduce the dimensionality of agents' preferences. The utility function assumptions and formulations used in the reviewed literature can be classified into three types. Table~\ref{tab:utility} summarises the formulations of these three types and lists the models that employ them. The underlying assumptions associated with the three types used in the literature are discussed below.

\begin{table} 
	\centering
    \caption{utility function types and assumptions used in the reviewed literature}
    \label{tab:utility}
    \begin{tabular}{p{0.12\linewidth} | p{0.25\linewidth}| p{0.15\linewidth}| p{0.25\linewidth}}
        & Assumptions & Examples & References\\ \hline
        
	\rowcolor{fancyBlue!25}
	Type I 
	& twice differentiable, monotonically increasing and strictly concave
	& quadratic, logarithmic
	& \cite{Cao2012, Barreto2013, Chen2014, Salfati2014, Namerikawa2015, Li2016marketpart1, Li2016marketpart2, Tavafoghi2017, Faqiry2017abudget, Zhou2018designing, TSAOUSOGLOU2019Near, Tsaousoglou2019personalized, Chen2019operating, Aizenberg2019, Samadi2011, Samadi2012, Okajima2014, Ma2014, Sinha2017, Yuan2017amechanism, VUELVAS2018limiting, MURAO2018real, Vuelvas2018acontract, Zhou2018contract, Zhong2018ADMM, STERIOTIS2018anovel, Barreto2015, Kang2017, BARRETO2019, Li2011, ma2017generalizing, Faqiry2019distributed, Hou2020reinforcement, Latifi2020arobust, Namerikawa2020distributed, Wang2020reward, wei2020mechanism}\rule{0pt}{2.6ex} \\
	
	Type II 
	& piecewise linear and concave
	& -
	& \cite{Bitar2013, Bitar2017, Muthirayan2019, Muthirayan2020Mechanism}\rule{0pt}{2.6ex} \\
	
	\rowcolor{fancyBlue!25}
	Type III 
	& constant utility if preferences are satisfied and zero otherwise
	& -
	& \cite{Mhanna2016afaithful, Mhanna2018afaithful}\rule{0pt}{2.6ex} \\
	
    \end{tabular}
\end{table}
\subsection{Type I: strictly concave, twice differentiable utility function}
The most commonly used formulation for the utility function is that it is strictly concave in consumption quantity. This formulation is based on the assumption that the utility function should satisfy the following properties:

\begin{enumerate}
    \item The utility function $U(q,\theta)$ is monotonically increasing in $q$, i.e.~more consumption yields higher utility.
    \begin{equation}
        \frac{\partial U(q,\theta)}{\partial q}\geq 0
    \end{equation}
    \item The marginal utility is nonincreasing in consumption quantity. This stems from the assumption that the larger the stock a person has the less value they would get from a given increase to that stock, i.e.~as their stock increases, their utility increases at a diminishing rate. The law of diminishing marginal utility is discussed at length in \cite{Ormazabal1995}. 
    \begin{equation}
        \frac{\partial^2 U(q,\theta)}{\partial q^2}\leq 0 
    \end{equation}
\end{enumerate}

Strictly concave functions satisfy the above properties. Examples of such functions include a bounded quadratic function, used in (\cite{Samadi2011, Samadi2012, Okajima2014, Ma2014, Sinha2017, Yuan2017amechanism, VUELVAS2018limiting, Vuelvas2018acontract, MURAO2018real, Zhou2018contract, Zhong2018ADMM, STERIOTIS2018anovel, Faqiry2019distributed, Latifi2020arobust, Wang2020reward}) and a logarithmic function (\cite{Barreto2015, Kang2017, BARRETO2019, Hou2020reinforcement}). Many of the reviewed papers model the utility function to be concave in quantity, however, this formulation has some limitations. One of the conditions that needs to hold for this formulation to be valid is that an agent's tastes and preferences do not change during the time-cycle of the mechanism. This can be challenging in the electricity market setting because most of the reviewed models propose day-ahead bidding and scheduling of electricity consumption. Additionally, the law of diminishing utility may not hold when dealing with task-related consumption, such as the case of electricity consumption. This is visible when an example of task related electricity consumption such as cooking is considered. Imagine that a meal would require 10 kWh to be done. consuming the first 9 kWh would have no value unless the last kWh is consumed. Therefore the utility rate of rise in such a scenario would be increasing with an increase in consumption rather than decreasing. This limitation is discussed at length in \cite{Ormazabal1995}. Additionally, one of the underlying assumptions for this formulation to hold is that electricity consumption is continuous. In reality, electricity consumption is task-related and is therefore a mixture of discrete and continuous energy demand as argued in \cite{Mhanna2014}. Furthermore, this type of valuation function does not take into account the intertemporal dependence of electricity consumption. This can lead to misrepresenting the preferences of an agent since they might change depending on past or future consumption \cite{Mhanna2016afaithful}.
\subsection{Type II: piecewise linear concave utility function}
The piecewise linear utility function is an extension of the concave utility function. The difference is that this type of utility functions is non-differentiable whereas strictly concave functions are twice differentiable. Indeed, piecewise linear utility functions are used in (\cite{Bitar2013, Bitar2017}) as a special case of concave utility functions so that the incentive compatibility property can hold in their proposed mechanism. The authors show that agents with non-constant marginal utility might gain (benefit) from being untruthful in reporting their private preferences. 
The authors in \cite{Muthirayan2019} also use a piecewise linear utility function to model the agent's satisfaction from electricity consumption. In their proposed mechanism, an agent is incentivised to report their baseline consumption and their constant marginal utility truthfully, upon which, the demand response aggregator (DRA) selects the agents that can achieve a demand reduction target in a cost efficient way.
\subsection{Type III: constant utility function}
The authors in \cite{Mhanna2016afaithful, Mhanna2018afaithful} argue that continuous concave utility functions cannot represent the task-related and combinatorial nature of electricity consumption. Instead, they use a discrete utility function that is constant if the reported consumer's preferences are satisfied and zero otherwise. In their proposed mechanism, household agents report their appliances' energy requirements and time flexibility. These preferences are then used to minimise the overall cost of energy. Although the intertemporal dependence of electricity consumption is captured in this mechanism, the dissatisfaction from shifting the operation of an appliance across the reported flexibility is not.

Despite types II and III being simple and computationally efficient, they fail to capture the complexity of consumption preferences. Consumers, therefore, might choose not to participate in the DSM schemes that force such utility formulations as they fail to represent their valuation of electricity consumption.

\section{Review of indirect mechanisms}
\label{sec:indirect}


Indirect mechanisms are capable of achieving desired social objectives without the need for a full revelation of the agents' private preferences. They become extremely important when these preferences are multi-dimensional and when market players are hesitant to participate because they are reluctant to reveal their private information. In these mechanisms, agents are required to provide incremental information that can indicate their private preferences. As a result, most indirect mechanisms are iterative in the sense that agents are required to update their bids at each step of the mechanism. An outcome is reached when a convergence criterion is met. Table~\ref{tab:class} lists the indirect mechanisms that were proposed for DSM in the literature. Some of the concerns that accompany indirect mechanisms are discussed below.

\subsubsection{Efficiency}
Since solving the optimal social welfare problem (stated in Eq.~\ref{eqn:SWO}) requires the full knowledge of the private utilities of the participating agents, indirect mechanisms usually suffer from efficiency losses because they only ask for partial information about these private utilities. In (\cite{Bitar2013, Bitar2017}) for instance, the authors propose a DSM mechanism where a supplier offers a non increasing deadline differentiated pricing bundle (i.e.~ lower price for demand with further deadline), to which customers respond by reporting their deadline differentiated demand. The optimal pricing bundle is then determined based on the supplier's optimal supply curve and the expected RE generation. The mechanism, however, makes an underlying assumption that customers are indifferent whether their demand is met before or at their reported deadline. This is likely not the case for most customers who would appreciate completing their tasks at the earliest possible. Because of this, losses in the welfare of users might arise, rendering the mechanism inefficient and unappealing in practice.


The authors in (\cite{Faqiry2017abudget, Faqiry2019double}) propose a double auction for energy trading between prosumers in a microgrid. The coordinator in this mechanism announces initial demand allocations to the consumers and initial supply requirements and price to the producers. Buyers then bid their optimal prices for the allocated demand and suppliers bid their optimal energy supply at the announced price. The coordinator then updates the market price and energy allocations by solving SWO. This is repeated until a convergence criterion is met. One of the limitations of this mechanism is that agents are assumed to be price taking, whose actions do not influence market price. In reality however, agents can be strategic and might know their actions' effect on the market. Another limitation of this mechanism is that final allocations are determined centrally. This can lead to efficiency losses if agents choose not to follow these allocations due to uncertainty in supply or demand.
A posted price mechanism for energy markets is presented in \cite{Tan2019posted}. The authors design optimal price profiles for each time interval that agents (arriving at different stages) can take or leave. Although the mechanism proves to be robust against uncertainties in demand and is privacy preserving, it suffers from losses in efficiency due to the partial knowledge of agents' preferences.
The authors in \cite{Hou2020reinforcement} propose a pricing mechanism for EV charging stations by employing a Markov decision process to model the strategic interactions between the station and its connected EVs. Although this pricing mechanism can maximise the long-term revenue of the charging station, it does not however maximise the social welfare of its users.
\subsubsection{Convergence}

Given that most of the indirect mechanisms proposed for DSM are iterative, evaluating the convergence speed of such mechanisms is essential to their successful implementation in practice. The authors in \cite{Majumder2015} propose a double auction mechanism for energy transactions between prosumers in a micro grid, where buyers and sellers bid their prices for each opposing agent. The micro grid controller then determine the demand and supply vectors that optimise the social welfare, based on which, agents update their bids in an iterative and distributed manner. Given that agents are required to solve a number of optimisation problems at every step of the mechanism, convergence to the optimal allocations might be slow, which would limit the proposed auction from being implemented in practice.


An RTP scheme for prosumers in a microgrid is proposed in \cite{Namerikawa2015, Namerikawa2020distributed}, where the authors use subsidies to align selfish behaviour with social behaviour. These subsidies are computed through an iterative power and price information exchange between the agents. A limitation of this mechanism is that convergence cannot be guaranteed unless the private utilities of the agents are known. 
The authors in \cite{TSAOUSOGLOU2019Near} propose a mechanism for strategic agents that can provide flexibility services to the grid. The authors use a reward/penalty billing rule that penalises demand that is highly correlated with aggregate demand rewards it when it is less correlated. In this mechanism, an iterative simulated annealing method is used to determine the agents' payments, where agents need to to solve a set of optimisation problems at each iteration. This method can be unpredictable and slow in terms of rate of convergence. In \cite{Tsaousoglou2019personalized}, the authors propose an RTP pricing scheme that is customised for each user, which offers lower prices for users who consume a less percentage of their desired level and higher prices for those who consume a higher percentage of their desired consumption level. Although social welfare can be maximised using this scheme, the prices are determined in an iterative manner where each agent reports their optimal consumption in response to the updated price at each step. This procedure needs to be performed for each agent, which can lead to high computational complexity and slow convergence.

A bilevel auction mechanism for demand response aggregators (DRA) is proposed in \cite{Faqiry2019distributed}. In this mechanism, DRAs bid for their energy share at the higher level, and prosumers bid for their supply/demand at the lower level. At each step of this auction, aggregators receive their allocated energy information from the utility company and implement their lower level auction by interacting with their selling and buying agents sequentially and separately. The complexity of this 2-level auction format along with the asymmetry in bidding between sellers and buyers in the lower level auction might cause convergence to be slow.

\subsubsection{Privacy}

Indirect mechanisms can overcome privacy concerns of consumers who do not wish to reveal their private valuation information fully by asking for partial incremental information about their private preferences. Nonetheless, in some of the indirect mechanisms that are iterative, the agents' updated reports might be exploited to reveal some of their private valuations and preferences. Moreover, some of the indirect mechanisms request their agents to report their desired consumption schedule given an announced price. This can be more unacceptable to some consumers than revealing their private valuation.
A clock-proxy auction mechanism for online scheduling of demand is proposed in \cite{Chapman2017aniterative}. In this 2-phase auction format, (ball park prices are agreed upon in the clock phase)  price discovery is implemented in the clock phase with iterative price adjustments and bid updates. Whereas the final uniform time-dependent prices and demand schedules are determined for each time interval in the proxy phase. Although this auction does not require agents to reveal their private information fully, a closer look into their bid updates can indicate an approximate representation of their utilities, thus compromising their privacy.
In (\cite{Zhou2018designing, BARRETO2019}), the authors propose incentive mechanisms that encourage social consumption behaviour. In both mechanisms, agents update their consumption schedules sequentially in response to an announced price or power signal. This might be limiting in practice as consumers would be reluctant to share their private consumption patterns. 


\subsubsection{Intermediate False Reporting}

One of the major concerns that need to be investigated before iterative mechanisms can be applied in practice is unfaithful reporting in intermediary steps. This refers to when experienced agents who have an estimate of the number of iterations required for convergence may submit false bids in the iterations before the final one if it is in their self-interest. This concern becomes evident when we consider the mechanisms proposed in (\cite{LiChen2011, Tushar2015, STERIOTIS2018anovel, Wang2020reward}), where an iterative exchange of price and power information is used to determine the optimal price and demand. The authors, however, do not investigate whether agents reporting their demand falsely in intermediary iterations can achieve higher utility gains than truthfully reporting their demand at every iteration.

\subsubsection{Consumption-Time Dependence}

Some of the proposed indirect mechanisms for DSM fail to model the temporal dependence of electricity consumption. In (\cite{Barreto2013, Barreto2015}) for instance, an incentive mechanism is proposed to modify the consumption behaviour of selfish customers. In this mechanism, customers' consumption schedules are used to compute their customised incentives. The authors, however, assume that the utility customers get from consuming energy at a given time does not depend on their consumption prior to that time or their planned consumption after it. This can be limiting since customers usually value their consumption differently depending on their previous and future planned consumption. 
The authors in \cite{Latifi2020arobust} propose a DSM mechanism where customers in a neighborhood cooperatively estimate the real-time price of the future time slot and then schedule their consumption accordingly to achieve their best self-interest. The regressors used in the estimation method, which are the historical demand profiles of the neighborhood agents, are assumed to be temporally independent. This underlying assumption may lead to inefficiencies due to the temporal dependence of electricity consumption. 


\section{Conclusion}
\label{sec:con}
In this paper, MD applications in market-based DSM were reviewed. the available literature were classified according to four design criteria; (1) direct or indirect revelation, (2) central or distributed allocation decision making, (3) iterative or one-shot bidding sequence, and (4) scalability of the proposed mechanisms. These design tools were investigated and their benifits/drawbacks in DSM applications were analysed. A challenging preliminary in many of the proposed MD-DSM approaches is the mathematical formulation of the utility function. Three types of this formulation were found in the literature within the scope of this review. The limitations and concerns of each type of formulation were discussed. As a promising subfield of MD applications in DSM, indirect revelation mechanisms were thoroughly reviewed and the concerns associated with these mechanisms were investigated.
One of the areas that requires further research is the design of indirect, non-iterative mechanisms for DSM that adopt a distributed allocation architecture and that is computationally simple. Another future research direction is the design of combinatorial auctions, which combine a number of goods and services into one item, for DSM. This can be promising because prosumers may value a combination of generation, consumption, storage and demand flexibility services differently when compared to each good or service separately. Due to privacy concerns, further research into DSM mechanisms that can offer optional revelation schemes for electricity consumers/prosumers may be required to attract wider customers' participation.

\ifCLASSOPTIONcaptionsoff
  \newpage
\fi

\end{document}